\documentclass{iau-JDSS}
\usepackage{graphicx}

\title[The Gaia Astrometric Survey] %% give here short title %%
{The Gaia Astrometric Survey}

\author[A. Sozzetti]   %% give here short author list %%
{Alessandro Sozzetti$^1$%
}

\affiliation{$^1$INAF - Osservatorio Astronomico di Torino, \break
Strada Osservatorio 20, 10025 Pino Torinese (TO), Italy \break email: sozzetti@oato.inaf.it%\\[\affilskip]
}

\pubyear{2010}
\volume{Volume 15}  %% insert here IAU Highlights of Astronomy Volume Number
\pagerange{??--??}
\date{?? and in revised form ??}
 \jname{Highlights of Astronomy, Volume 15}
\editors{Ian F Corbett, ed.}
\begin{document}

\maketitle

\begin{abstract}

In its all-sky survey, the ESA global astrometry mission Gaia will
perform high-precision astrometry and photometry for 1 billion
stars down to $V = 20$ mag. The data collected in the Gaia
catalogue, to be published by the end of the next decade, will
likely revolutionize our understanding of many aspects of stellar
and Galactic astrophysics. One of the relevant areas in which the
Gaia observations will have great impact is the astrophysics of
planetary systems. This summary focuses on a) the complex
technical problems related to and challenges inherent in correctly
modelling the signals of planetary systems present in measurements
collected with a space-borne observatory poised to carry out
precision astrometry at the micro-arcsecond ($\mu$as) level, and
b) on the potential of Gaia $\mu$as astrometry for important
contributions to the astrophysics of planetary systems.

%% add here a maximum of 10 keywords, to be taken form the file <Keywords.txt>
\end{abstract}

\firstsection % if your document starts with a section,
              % remove some space above using this command.
\section{Introduction}\label{intro}

The Gaia all-sky survey, due to launch in Spring 2012, will
monitor astrometrically, during its 5-yr nominal mission lifetime,
all point sources (stars, asteroids, quasars, extragalactic
supernovae, etc.) in the visual magnitude range $6-20$ mag, a huge
database encompassing $\sim10^9$ objects. Using the continuous
scanning principle first adopted for Hipparcos, Gaia will
determine the five basic astrometric parameters (two positional
coordinates $\alpha$ and $\delta$, two proper motion components
$\mu_\alpha$ and $\mu_\delta$, and the parallax $\varpi$) for all
objects, with end-of-mission precision between 6 $\mu$as (at $V=6$
mag) and 200 $\mu$as (at $V=20$ mag).

Gaia astrometry, complemented by on-board spectrophotometry and
(partial) radial velocity information, will have the precision
necessary to quantify the early formation, and subsequent
dynamical, chemical and star formation evolution of the Milky Way
Galaxy. The broad range of crucial issues in astrophysics that can
be addressed by the wealth of the Gaia data is summarized by e.g.,
Perryman et al. (2001). One of the relevant areas in which the
Gaia observations will have great impact is the astrophysics of
planetary systems (e.g., Casertano et al. 2008), in particular
when seen as a complement to other techniques for planet detection
and characterization (e.g., Sozzetti 2009).

\section{Astrometric Modelling of Planetary Systems}\label{model}

The problem of the correct determination of the astrometric orbits
of planetary systems using Gaia data (highly non-linear orbital
fitting procedures, with a large number of model parameters) will
present many difficulties. For example, it will be necessary to
assess the relative robustness and reliability of different
procedures for orbital fits, together with a detailed
understanding of the statistical properties of the uncertainties
associated with the model parameters. For multiple systems, a
trade-off will have to be found between accuracy in the
determination of the mutual inclination angles between pairs of
planetary orbits, single-measurement precision and redundancy in
the number of observations with respect to the number of estimated
model parameters. It will constitute a challenge to correctly
identify signals with amplitude close to the measurement
uncertainties, particularly in the presence of larger signals
induced by other companions and/or sources of astrophysical noise
of comparable magnitude. Finally, in cases of multiple-component
systems where dynamical interactions are important (a situation
experienced already by radial-velocity surveys), fully dynamical
(Newtonian) fits involving an n-body code might have to be used to
properly model the Gaia astrometric data and to ensure the short-
and long-term stability of the solution (see Sozzetti 2005).

All the above issues could have a significant impact on Gaia's
capability to detect and characterize planetary systems. For these
reasons, within the pipeline of Coordination Unit 4 (object
processing) of the Gaia Data Processing and Analysis Consortium
(DPAC), in charge of the scientific processing of the Gaia data
and production of the final Gaia catalogue to be released sometime
in 2020, a Development Unit (DU) has been specifically devoted to
the modelling of the astrometric signals produced by planetary
systems. The DU is composed of several tasks, which implement
multiple robust procedures for (single and multiple) astrometric
orbit fitting (such as Markov Chain Monte Carlo and genetic
algorithms) and the determination of the degree of dynamical
stability of multiple-component systems.

\section{The Legacy of Gaia}\label{gaia}

Using Galaxy models, our current knowledge of exoplanet
frequencies, and Gaia's estimated precision ($\sim 10$ $\mu$as) on
bright targets ($V < 13$), Casertano et al. (2008) have shown how
Gaia's main strength will be its ability to measure
astrometrically actual masses and orbital parameters for possibly
thousands of giant planets, and to determine the degree of
coplanarity in possibly hundreds of multiple-planet systems. Its
useful horizon for planet detection (encompassing
$\sim3\times10^5$ stars) extends as far as the nearest
star-forming regions (e.g., Taurus at $d\simeq 140$ pc) for
systems with massive giant planets ($M_p \gtrsim 2-3$
$M_\mathrm{J}$) on $1<a<4$ AU orbits around solar-type hosts, and
out to $d\sim 30$ pc for Saturn-mass planets with similar orbital
semi-major axes around late-type stars.

In summary, Gaia holds promise for crucial contributions to many
aspects of planetary systems astrophysics, in combination with
present-day and future extrasolar planet search programs. For
example, the Gaia data, over the next decade, will allow us to a)
significantly refine of our understanding of the statistical
properties of extrasolar planets, b) carry out crucial tests of
theoretical models of gas giant planet formation and migration, c)
achieve key improvements in our comprehension of important aspects
of the formation and dynamical evolution of multiple-planet
systems, d) provide important contributions to the understanding
of direct detections of giant extrasolar planets, and e) collect
essential supplementary information for the optimization of the
target lists of future observatories aiming at the direct
detection and spectroscopic characterization of terrestrial,
habitable planets in the vicinity of the Sun.

%\begin{acknowledgments}
%\end{acknowledgments}

\end{document}